\documentclass[letterpaper,11pt]{article}
\usepackage{epsfig}

\def\be{\begin{equation}}
\def\ee{\end{equation}}
\def\bc{\begin{center}}
\def\ec{\end{center}}
\def\bea{\begin{eqnarray}}
\def\eea{\end{eqnarray}}

\def\nn{\nonumber}

\def\EP{effective potential}
\def\vol{V_{1\ell}}
\def\vtl{V_{2\ell}}

\def\sigu{\Sigma_1}
\def\sigd{\Sigma_2}
\def\eps{\epsilon}
\def\drbar{\overline{\rm DR}}
\def\drsmall{\overline{\rm \scriptscriptstyle DR}}
\def\lnb{\overline{\ln}}
\def\sq2{\sqrt{2}}
\def\hlf{\frac{1}{2}}
\def\s2t{s_{2\theta}}
\def\C2t{c_{2\theta}}
\def\ths{\theta_{\tilde{t}}}

\def\at{\alpha_t}
\def\ab{\alpha_b}
\def\as{\alpha_s}
\def\oat{{\cal O}(\at)}

\def\oatas{{\cal O}(\at\as)}
\def\oabas{{\cal O}(\ab\as)}
\def\oabqatab{{\cal O}(\ab^2 + \at\ab)}
\def\oatasatq{{\cal O}(\at\as + \at^2)}
\def\oatasabas{{\cal O}(\at\as +\ab\as)}
\def\oatasabasatq{{\cal O}(\at\as +\ab\as + \at^2)}
\def\oatq{{\cal O}(\at^2)}

\def\mgl{m_{\tilde{g}}}
\def\mt{m_t}
\def\mh{m_h}
\def\mz{m_{\scriptscriptstyle Z}}
\def\polemz{M_{\scriptscriptstyle Z}}
\def\ma{m_{\scriptscriptstyle A}}

\def\mgut{M_{\scriptscriptstyle GUT}}
\def\msqu{m_{\tilde{t}_1}^2}
\def\msqd{m_{\tilde{t}_2}^2}
\def\diff{\msqu-\msqd}
\def\m0{m_0}
\def\mhf{m_{1/2}}
\def\qmin{Q_{\rm min}}
\def\qbest{Q_{*}}
\def\gev{{\rm GeV}}

\def\mylg{\ln\frac{\mgl^2}{Q^2}}
\def\mylt{\ln\frac{m_t^2}{Q^2}}
\def\myltu{\ln\frac{\msqu}{Q^2}}
\def\myltd{\ln\frac{\msqd}{Q^2}}

\def\myltq{\ln^2\frac{m_t^2}{Q^2}}
\def\myltuq{\ln^2\frac{\msqu}{Q^2}}

\newcommand{\gsim}{\lower.7ex\hbox{$\;\stackrel{\textstyle>}{\sim}\;$}}
\newcommand{\lsim}{\lower.7ex\hbox{$\;\stackrel{\textstyle<}{\sim}\;$}}

\newenvironment{appendletterA}
 {
  \setcounter{section}{0}
  \setcounter{equation}{0}
  
 }{
 }
\newenvironment{appendletterB}
 {
  \setcounter{equation}{0}
  
 }{
 }

\begin{document}

\thispagestyle{empty}

\bc
\hfill{MPI-PhT/2002-77} \\
\hfill{TUM-HEP-495/02}  \\
\ec

\vspace{1.7cm}
\bc
{\LARGE\bf Two--loop corrections to Radiative Electroweak   } 
\ec

\bc
{\LARGE\bf Symmetry Breaking in the MSSM}
\ec

\vspace{1.4cm}

\bc

{\Large \sc Athanasios~Dedes~$^{a,}$\footnote{{\tt
      dedes@ph.tum.de}},
 ~Pietro~Slavich~$^{b,\,c,}$\footnote{\tt slavich@mppmu.mpg.de}}\\
\vspace{1.2cm}

${}^a$
{\em Physik Department, Technische Universit\"at M\"unchen,\\
D--85748 Garching, Germany}
\vspace{.3cm}

${}^b$
{\em Institut f\"ur Theoretische Physik, Universit\"at Karlsruhe,\\
Kaiserstrasse 12, Physikhochhaus, D--76128 Karlsruhe, Germany}
\vspace{.3cm}

${}^c$
{\em Max Planck Institut f\"ur Physik,\\
F\"ohringer Ring 6, D--80805 M\"unchen, Germany}

\ec

\vspace{0.8cm}

\centerline{\bf Abstract}
\vspace{2 mm}
\begin{quote} \small
We study the $\oatasatq$ two--loop corrections to the minimization
conditions of the MSSM effective potential, providing compact
analytical formulae for the Higgs tadpoles. We connect these results
with the renormalization group running of the MSSM parameters from the
grand unification scale down to the weak scale, and discuss the
corrections to the Higgs mixing parameter $\mu$ and to the running
CP--odd Higgs mass $\ma$ in various scenarios of gravity--mediated
SUSY breaking. We find that the $\oatas$ and $\oatq$ contributions
partially cancel each other in the minimization conditions.  In
comparison with the full one--loop corrections, the $\oatasatq$
two--loop corrections significantly weaken the dependence of the
parameters $\mu$ and $\ma$ on the renormalization scale at which the
\EP\ is minimized.  The residual two--loop and higher--order
corrections to $\mu$ and $\ma$ are estimated to be at most 1\% in the
considered scenarios.
\end{quote}
\vfill
\newpage
\setcounter{equation}{0}
\setcounter{footnote}{0}
\vskip2truecm


\section{Introduction}


One of the most attractive features of the Minimal Supersymmetric
extension of the Standard Model (MSSM)~\cite{Nilles}, is the fact that
it provides a mechanism for breaking radiatively the electroweak gauge
$SU(2)_L \times U(1)_Y$ symmetry down to $U(1)_{\rm EM}$. It was first
shown~\cite{Ross} that a supersymmetry (SUSY) breaking term for the
gluino can induce an effective potential which spontaneously breaks
the electroweak symmetry. At the same time, a mechanism relying on the
renormalization group evolution from a grand unification (GUT) scale
$\mgut$ down to the weak scale was proposed~\cite{Inoue}. In this
framework, at the scale $\mgut$, the parameters entering the scalar
potential of the MSSM obey simple boundary conditions dictated by the
underlying theory of SUSY breaking, and the electroweak symmetry is
unbroken. When the parameters are evolved down to the weak scale by
means of the MSSM renormalization group equations (RGE), which amounts
to resumming the leading logarithmic corrections to all orders, the
soft SUSY--breaking mass $m^2_{H_2}$ is driven towards negative
values, due to corrections controlled by the top Yukawa coupling
$h_t$. This helps to destabilize the origin in field space, so that
the Higgs fields acquire non--vanishing vacuum expectation values
(VEVs) and the electroweak symmetry is spontaneously broken. Although
the studies in Refs.~\cite{Ross,Inoue} where differing on the initial
boundary conditions, the result was one: the radiative electroweak
symmetry breaking (REWSB) takes place if the top Yukawa coupling is
large, such that $60 \; \gev \lsim m_t \lsim 200 \; \gev$, with the upper bound
coming from the requirement that $h_t$ remains in the perturbative
range up to the GUT scale. It could be a coincidence that the top quark
is found at the Tevatron to have mass around 175 GeV, but certainly
this is consistent with the REWSB mechanism in the MSSM.

In the RGE--improved potential of the MSSM employed at tree level, the
VEVs of the Higgs fields, and the occurrence of spontaneous symmetry
breaking itself, depend critically on the renormalization scale at
which the parameters entering the potential are computed; an
inappropriate choice of that scale can lead to results that are even
qualitatively wrong. In fact, the electroweak symmetry is either
broken or unbroken, independently of the renormalization scale choice,
and the critical behavior described above is just an artifact of the
tree--level approximation. The correct way of determining the ground
state of the theory is to minimize the Coleman--Weinberg effective
potential \cite{Coleman}, i.e. the tree--level potential plus a
correction coming from the sum of all the one--loop diagrams with
zero--momentum external lines. Since Refs.~\cite{Kounnas,grz} this
procedure has become standard in the renormalization group analyses of
the MSSM (for early examples see Refs.~\cite{rgemssm,studies}).

The effective potential is also a useful tool for computing the
leading corrections to the MSSM Higgs masses, both at the one
loop~\cite{higgs1l} and the two loop~\cite{hh,ez,noi,noibot,marthiggs}
level\footnote{Other two--loop computations of the MSSM Higgs masses
have been performed in the renormalization group~\cite{rgehiggs} and
diagrammatic \cite{diaghiggs} approaches.}, in the approximation of
zero external momentum.  The leading one--loop corrections are $\oat$,
i.e. they are controlled by the top Yukawa coupling $h_t \equiv
\sqrt{4 \pi \at}$. For stop masses of ${\cal O}(1\, {\rm TeV})$, such
corrections increase by 40--60 GeV the mass $\mh$ of the lightest Higgs
boson (which at tree level must be lighter than $\mz$), allowing it to
escape the direct searches at LEP. Also, the leading two--loop
corrections have sizeable effects: the $\oatas$ corrections,
controlled by the strong gauge coupling $g_s \equiv \sqrt{4 \pi
\as}\,$, typically reduce $\mh$ by 15--20 GeV, whereas the $\oatq$
ones may increase it by up to 7--8 GeV.

Motivated by the relevance of the $\oatasatq$ two--loop corrections in
the case of the Higgs masses, we study in this paper the effect of the
same corrections on the electroweak symmetry breaking conditions. The
contributions to the two--loop MSSM effective potential that are
relevant to the $\oatasatq$ corrections have been discussed in
Refs.~\cite{hh,ez,noi}, and a complete computation of the two--loop
effective potential has been presented in
Ref.~\cite{martinEP}. However, practical studies of REWSB usually
require explicit formulae for the Higgs tadpole diagrams, i.e. the
first derivatives of the effective potential with respect to the Higgs
fields. Such formulae are presently available at the one--loop order
\cite{rgemssm}, but they have not been presented so far at the
two--loop order.  Using the techniques developed in Ref.~\cite{noi},
we compute in this paper explicit and compact analytical expressions
for the two--loop $\oatasatq$ part of the tadpoles. As a byproduct
from our $\oatas$ corrections we obtain also the $\oabas$ corrections,
that are relevant for large values of $\tan\beta$. Once we assume that
the electroweak symmetry is indeed broken, giving rise to the observed
value of the $Z$ boson mass, the corrections to the tadpoles translate
into $\oatasatq$ corrections to the values of $\mu$, the Higgs mass
term in the superpotential, and $\ma$, the running mass of the $A$
boson.  We discuss the effect of our two--loop corrections in the
framework of gravity mediated SUSY breaking \cite{mSUGRA}, also
denoted as minimal supergravity (mSUGRA), referring in particular to
various ``benchmark'' scenarios suggested at
Snowmass~\cite{snowmass}. We find that the inclusion of the
$\oatasatq$ corrections significantly improves the renormalization
scale dependence of the results,  and that partial cancellations
occur between the $\oatas$ corrections and the $\oatq$ ones. Our
corrections are also required for consistency in the $\oatasatq$
two--loop computation of the MSSM Higgs masses, if the input
parameters are computed via renormalization group evolution from a set
of high energy boundary conditions.

The paper is organized as follows: in section 2 we recall the basic
concepts of radiative electroweak symmetry breaking, and introduce
some notation which will be used in the rest of the paper; in section
3 we describe the main features of our $\oatasatq$ computation of the
two--loop tadpoles; in section 4 we discuss the numerical effect of
our corrections, and we show how they improve the dependence of $\mu$
and $\ma$ on the renormalization scale at which the effective
potential is minimized; section 5 contains our conclusions. In
addition, we present in the appendix A some useful formulae for the
integrals entering the two--loop effective potential, and in the
appendix B the explicit analytical formulae for the $\oatas$ part of
the corrections. The formulae for the $\oatq$ part are indeed rather
long, thus we make them available, upon request, in the form of a
computer code~\footnote{E--mail: {\tt slavich@mppmu.mpg.de}}.


\section{Radiative electroweak symmetry breaking}
\label{sec:general}

We start our discussion from the tree--level scalar potential of the
MSSM, that reads, keeping only the dependence on the neutral Higgs
fields $H_1^0$ and $H_2^0$:
\be
\label{V0}
V_0  =  \Lambda +
m_1^2 \, \left| H_1^0 \right|^2 
+ m_2^2 \, \left| H_2^0 \right|^2 
+ m_3^2 \, \left( H_1^0 H_2^0 + {\rm h.c.} \right)
+ {g^2 +g^{\prime\,2} \over 8} \left(
|H_1^0|^2 - |H_2^0|^2 \right)^2 \, ,
\ee
where: $\Lambda$ is a field--independent vacuum energy; $m_1^2 =
m_{H_1}^2 + \mu^2$, $\,m_2^2 = m_{H_2}^2 + \mu^2$ (we assume $\mu$ to be
real, neglecting all possible CP--violating phases); $m_{H_1}^2$,
$m_{H_2}^2$ and $m_3^2$ are soft SUSY--breaking masses; $g$ and $g'$
are the $SU(2)_L$ and $U(1)_Y$ gauge couplings, respectively.  At the
classical level, the mass parameters entering $V_0$ must satisfy the
following conditions:
\be
\label{conds}
m_1^2+m_2^2 \;\geq\; 2\,|m_3^2| \;\;,\hspace{2cm}
m_1^2\,m_2^2 \;\leq\; m_3^4\,.
\ee
The first condition guarantees that the potential is bounded from
below; the second condition destabilizes the origin in field space,
making sure that the neutral components of the Higgs fields acquire
non--vanishing VEVs $\langle H_1^0 \rangle \equiv v_1/\sq2$ and
$\langle H_2^0 \rangle \equiv v_2/\sq2$.  It is not restrictive to
choose $m_3^2$ real and negative\footnote{Our conventions differ 
by a sign in the parameters $\mu$ and $m_3^2$ 
with respect to those used in the second paper of Ref.\cite{Nilles}.}, 
so that $v_1$ and $v_2$ are real and
positive, and the neutral Higgs fields can be decomposed into their
VEVs plus their CP--even and CP--odd fluctuations as $H_i^0 = (v_i +
S_i + i P_i)/\sq2$.

Since the parameters entering $V_0$ are taken as "running" ones (i.e.,
they vary with the renormalization scale), also the validity of the
conditions in Eq.~(\ref{conds}) depends on the scale, as well as the
numerical values of $v_1$ and $v_2$. As discussed in Ref.~\cite{grz},
the minimization of the tree--level potential may lead to grossly
inaccurate results, unless the renormalization scale is chosen in such
a way that the radiative corrections to the scalar potential are
small. To obtain the correct results, one should rather minimize the
effective potential $V_{\rm eff}$, defined as:
\be
\label{effp}
V_{\rm eff} = V_0 + \Delta V \, ,
\ee
where $\Delta V$ contains the radiative corrections to the scalar potential
$V_0$. The minimization conditions for $V_{\rm eff}$ can be written
as:
\bea
\label{min1}
\frac{1}{v_1} \left.\frac{\partial V_{\rm eff}}{\partial S_1}
\right|_{\rm min} &=&
m^2_{H_1} + \mu^2  + \frac{g^2 + g^{\prime \,2}}{4}\,(v_1^2-v_2^2)
+ m_3^2 \,\frac{v_2}{v_1} + \sigu \;\;\; = \;0\,,\\
&&\nn\\
\label{min2}
\frac{1}{v_2} \left.\frac{\partial V_{\rm eff}}{\partial S_2}
\right|_{\rm min} &=&
m^2_{H_2} + \mu^2 + \frac{g^2 + g^{\prime \,2}}{4}\,(v_2^2-v_1^2)
+ m_3^2 \,\frac{v_1}{v_2} + \sigd \;\;\; = \;0\,,
\eea
where the ``tadpoles'' $\sigu$ and $\sigd$ are defined as:
\be
\label{sigma}
\Sigma_i \equiv \frac{1}{v_i} \left.\frac{\partial \Delta V}{\partial S_i}
\right|_{\rm min}\,.
\ee
In principle, a renormalization group study of the MSSM should start
from some large scale $\mgut$, where the input parameters have a
simple structure dictated by the underlying theory of SUSY
breaking, and the electroweak symmetry is unbroken. The parameters are
then evolved, by means of appropriate renormalization group equations,
down to some lower scale, where the electroweak symmetry breaking
occurs and the VEVs $v_1$ and $v_2$ can be obtained by solving
Eqs.~(\ref{min1})--(\ref{min2}).  A set of high--energy input
parameters is then acceptable if it leads to the correct value of the
squared running mass for the $Z$ boson, $\mz^2 = (g^2 + g^{\prime
\,2})\,(v_1^2+v_2^2)/4$.  However, in most practical applications of
the renormalization group procedure, it is more convenient to assume
that {\em there is} successful electroweak symmetry breaking, and
trade two of the high--energy input parameters for $v_1$ and $v_2$
(or, equivalently, for $v^2 \equiv v_1^2 + v_2^2$ and $\tan\beta
\equiv v_2/v_1$).  Eqs.~(\ref{min1})--(\ref{min2}) can thus be
rephrased into the following conditions among the parameters at the
weak scale:
\bea
\label{eqmu}
\mu^2 & = & - \frac{\mz^2}{2} 
+ \frac{m^2_{H_1} + \sigu - (m^2_{H_2} + \sigd) 
\,\tan^2\beta}{\tan^2\beta - 1}\,,\\
&\nn\\
\label{eqm3}
m_3^2 & = & -\hlf\,\sin 2\beta\, \left( m^2_{H_1} + m^2_{H_2} 
+ 2\,\mu^2 + \sigu + \sigd \right) \, ,
\eea

\noindent
i.e., the terms proportional to $\sigu$ and $\sigd$ in the above
equations can be viewed as the radiative corrections to the values of
$\mu^2$ and $m_3^2$ obtained from the requirement of successful
electroweak symmetry breaking. We recall that $m_3^2$ is related to
the squared running mass for the $A$ boson through $\ma^2 = -
2\,m_3^2/ \sin 2\beta$. Notice also that the sign of $\mu$ is not
fixed by Eq.~(\ref{eqmu}), and it must be supplemented as an
additional input quantity.  If the right side of Eq.~(\ref{eqmu}) is
such that $\mu^2$ is negative, then our choice of input parameters is
inconsistent, and the electroweak symmetry fails to be broken.
We remark in passing that the choice of the input parameters is
constrained by further requirements: it must lead to a spectrum of
physical masses for the MSSM superpartners and Higgs bosons compatible
with the present experimental lower bounds, and such that the lightest
supersymmetric particle (LSP) is electrically neutral; it must satisfy
phenomenological constraints coming from radiative B-meson decays, muon
anomalous magnetic moment and cosmological relic density; finally, it
must guarantee that the MSSM scalar potential is bounded from below
and does not lead to charge and color breaking minima. However, a
detailed study of the (theoretically and experimentally) allowed
regions in the MSSM parameter space goes beyond the scope of this
paper, and will not be pursued in the following.

The full one--loop corrections to the REWSB conditions have been
extensively discussed in the literature~\cite{rgemssm} in the
framework of the mSUGRA scenario. The dominant one--loop contributions to
$\sigu$ and $\sigd$ come to a large extent from the top/stop (and, for
large $\tan\beta$, bottom/sbottom and tau/stau) diagrams. The
contributions of the diagrams involving charginos and neutralinos can
also be sizeable and  comparable to the top/stop ones
in some regions of the parameter space, while the
Higgs and gauge bosons and the first two generations of (s)quarks and
(s)leptons give only subdominant corrections.

In the following sections we provide explicit analytical formulae for
the two--loop top/stop contributions to the tadpoles $\sigu$ and
$\sigd$, resulting into $\oatasatq$ corrections to $\mu^2$ and to the
running mass $\ma^2$. In analogy with the case of the Higgs masses, we
expect such corrections to be the leading ones at the two--loop level,
giving rise to sizeable effects at least in some regions of the MSSM
parameter space. In addition, most public codes that compute the MSSM
mass spectrum from a set of unified parameters at the scale $\mgut$,
such as {\em SuSpect} \cite{suspect}, {\em SoftSusy} \cite{softsusy},
{\em SPheno} \cite{spheno} and {\em FeynSSG} \cite{FeynSSG}, include a
two--loop $\oatasatq$ computation of the Higgs
masses,\footnote{Another widely used public code, {\em Isajet
7.58}~\cite{Isajet}, relies on a one--loop \EP\ computation of the
Higgs masses.} but employ one--loop results for the tadpoles (see
also Ref.~\cite{kraml} for a recent discussion). Since $\ma^2$ enters
the tree--level mass matrix of the CP--even Higgs bosons, the
$\oatasatq$ corrections to $\ma^2$ should be included in those codes
for consistency.


\section{Computation of the $\oatasatq$ corrections}
\label{sec:twoloop}

We shall now describe our two--loop, $\oatasatq$ computation of the
tadpoles $\sigu$ and $\sigd$, involving the first derivatives of
$\Delta V$ with respect to the CP--even parts of the neutral Higgs
fields [see Eq.~(\ref{sigma})]. The computation is consistently
performed by setting to zero all the gauge couplings but $g_s$ and by
keeping $h_t$ as the only non--vanishing Yukawa coupling (with a
slight abuse of language, in the following we will refer to this
approximation as to the gaugeless limit). In this limit, the
tree--level (field--dependent) spectrum of the MSSM simplifies
considerably: gauginos and Higgsinos do not mix; the charged and
neutral Higgsinos combine into Dirac spinors $\tilde{h}^0$ and
$\tilde{h}^\pm$ with degenerate mass $\mu$; the gaugino masses
coincide with the soft SUSY--breaking parameters $M_A\; (A =
1,2,3)\;$ (among them, only the gluino mass $\mgl = M_3$ is relevant
to our calculation); the only massive Standard Model (SM) fermion is
the top quark; all other fermions and gauge bosons have vanishing
masses; besides the top squarks, the only sfermion with
non--vanishing couplings is the bottom squark $\tilde{b}_L$; the
lighter CP--even Higgs boson, $h$, is massless, and the same is true
for the Goldstone bosons $G$ and $G^{\pm}$; all the remaining Higgs
states, $(H,A, H^\pm)$, have degenerate mass eigenvalues $\ma^2$. The
tree--level mixing angle in the CP--even sector is just
$\alpha=\beta-\pi/2$.

To begin with, we address the renormalization of the \EP. In the loop
expansion, the correction to the \EP\ can be decomposed as $\Delta V =
\vol + \vtl + \cdots$, where the ellipsis stand for higher
loops. Using the Landau gauge and dimensional reduction \cite{dred,drbp} in
$d = 4 - 2\eps$ dimensions, and including for later convenience also
terms that vanish when $\eps\rightarrow 0$, the unrenormalized
one--loop \EP\ reads:
\be
\label{v1loop}
\vol =
\frac{-1\;\;\;}{64 \pi^2}\,{\rm Str}\;{\cal M}^4 \left[\frac{1}{\eps}
+ \frac{3}{2} - \ln\frac{{\cal M}^2}{Q^2}  +\eps\,\left(
\frac{7}{4} - \frac{3}{2}\,\ln\frac{{\cal M}^2}{Q^2}
+\hlf \ln^2\frac{{\cal M}^2}{Q^2} + \frac{\,\pi^2}{12}\right)\,\right]\,,
\ee

\noindent
where ${\cal M}^2$ is the matrix of the field--dependent squared masses, 
and the supertrace of a generic function $f({\cal M}^2)$ is defined as 
a sum over the eigenvalues $m^2_i$: 
\be {\rm Str}\, f({\cal M}^2) = \sum_i (-1)^{2 s_i}(2 s_i+1)\,f(m^2_i)
\;,
\ee where $s_i$ is the spin of the corresponding particles. In
Eq.~(\ref{v1loop}), $Q^2 = 4\pi \mu^2 e^{-\gamma_E}$, i.e.  the finite
terms that are removed together with $1/\eps$ in the modified
subtraction schemes have been reabsorbed in the renormalization scale
(the same convention will be adopted in the following). In the
gaugeless limit described above, only the top and stop contributions
to $\vol$ are relevant, giving rise to $\oat$ contributions to $\sigu$
and $\sigd$.

\begin{figure}[p]
\begin{center}
\epsfig{figure=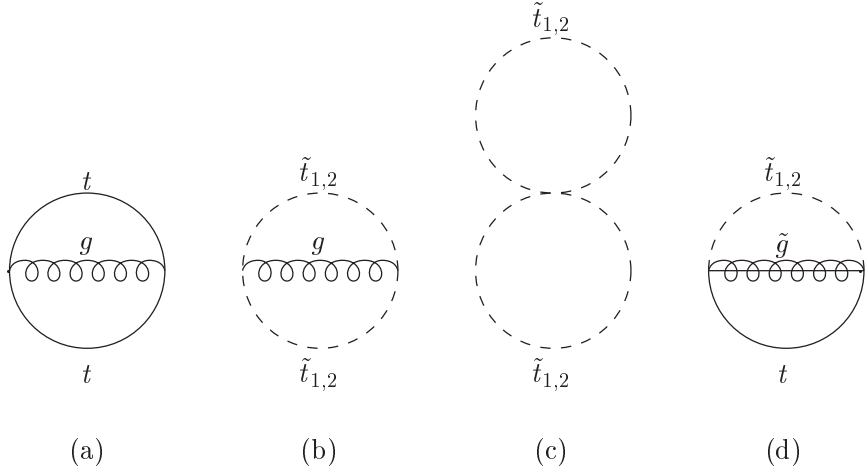,width=11.5cm}
\end{center}
\caption{Feynman diagrams that contribute to the two--loop \EP\ and
affect the $\oatas$ corrections to the electroweak symmetry breaking
conditions. The diagrams relevant to the $\oabas$ corrections can be
obtained with the replacement $t\to b$.}
\label{diagstr}
\end{figure}

\begin{figure}[p]
\begin{center}
\epsfig{figure=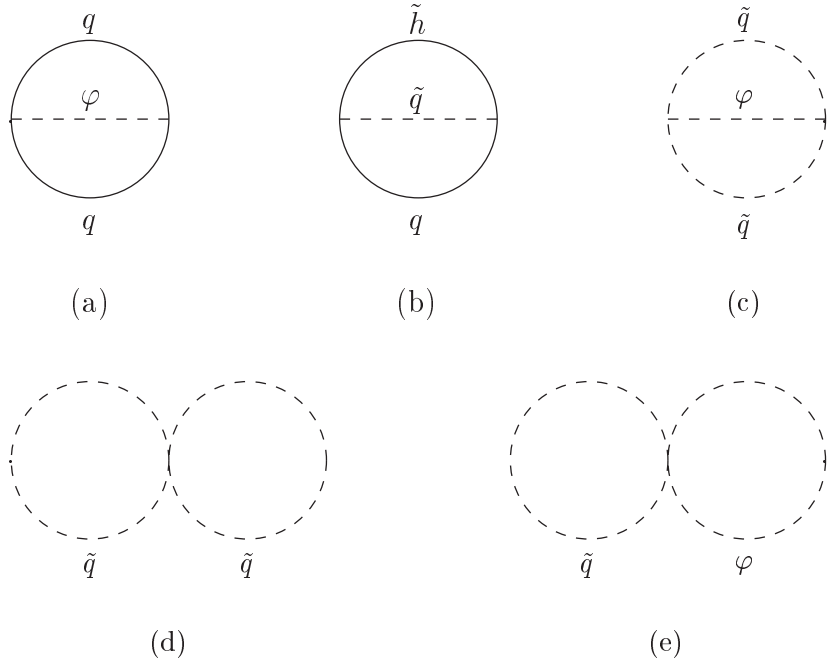,width=11cm}
\end{center}
\caption{The classes of Feynman diagrams that 
contribute to the two--loop \EP\ and affect the $\oatq$ 
corrections to the electroweak symmetry breaking conditions
[$q=(t,b)$, $\varphi = (H, h, G, A, H^\pm, G^\pm)$, $\tilde{h} =
(\tilde{h}^0, \tilde{h}^\pm)$, $\tilde{q} = (\tilde{t}_1,
\tilde{t}_2, \tilde{b}_L)$].}
\label{diagyuk}
\end{figure}

The Feynman diagrams that contribute to the two--loop \EP\ $\vtl$ and
give rise to $\oatas$ contributions to $\sigu$ and $\sigd$ are shown
in Fig.~\ref{diagstr}, while the diagrams relevant to the $\oatq$
contributions are shown in Fig.~\ref{diagyuk}. The corresponding
analytical formulae for $\vtl$ can be found e.g. in the last paper of
Ref.~\cite{ez}~\footnote{The formulae of Ref.~\cite{ez} are obtained
for vanishing CP--odd fields. While requiring some modifications
\cite{noi} for the computation of the CP--even Higgs boson masses in
terms of the physical $\ma$, those formulae can be used as they stand
for the purposes of the present analysis.}. These formulae involve two
basic integrals, $I(m_1^2,m_2^2,m_3^2)$ and $J(m_1^2,m_2^2)$, that
have been evaluated with different methods in
Refs.~\cite{jackjones,davtausk}.  Explicit expressions for $I$ and $J$
in the formalism of Ref.~\cite{davtausk} are presented in the appendix
A.

To carry out the renormalization of $V_{\rm eff}$, at the two--loop
order and in the $\drbar$ scheme \cite{dred,drbp}, we start from the
unrenormalized \EP, written in terms of generic bare parameters $x_i$.
Then, we expand the parameters as $x_i = x_i^{\drsmall} + \delta x_i$,
where $\delta x_i$ are purely divergent quantities, so that all the
poles in $1/\epsilon$ and $1/\epsilon^2$ are cancelled. After taking
the limit $\eps \rightarrow 0$, the non vanishing part of the
renormalized effective potential is:
\be
\label{effpR}
V_{\rm eff} = V_0(x^{\drsmall}_i) + V^{(1)}_{1\ell}(x^{\drsmall}_i)
+ V_{2\ell}^{(1)} +\frac{\partial V_{1\ell}^{(\epsilon)}}{\partial x_i}\,
\delta^{1\ell} x_i \, , 
\ee 
where $V_{1\ell}^{(1)}$ and $V_{2\ell}^{(1)}$ denote the finite parts
of the one--loop and two--loop \EP, respectively,
$V_{1\ell}^{(\epsilon)}$ denotes the terms proportional to $\eps$ in
Eq.~(\ref{v1loop}), and $\delta^{1\ell} x_i$ is the coefficient of
$1/\eps$ in the one--loop part of the generic counterterm (notice that
we need to compute explicitly only the one--loop counterterms for the
top and stop masses).  In Eq.~(\ref{effpR}), $V_0$ and
$V_{1\ell}^{(1)}$ are expressed in terms of $\drbar$--renormalized
parameters, while the renormalization of the parameters entering the
two--loop part is irrelevant, amounting to a higher--order (i.e.,
three--loop) effect. In summary, it is possible to define the
renormalized two--loop effective potential as:
\be
\label{tworen}
\widehat{V}_{2\ell} = V_{2\ell}^{(1)} 
+\frac{\partial V_{1\ell}^{(\epsilon)}}{\partial x_i} \,\delta^{1\ell} x_i\, . 
\ee
We have checked that $\widehat{V}_{2\ell}$ corresponds to the finite
part of the potential obtained by replacing the integrals $I$ and $J$
in $V_{2\ell}$ with the ``subtracted'' integrals $\hat{I}$ and
$\hat{J}$, first introduced in Ref.~\cite{jackjones}.  More precisely,
this is true only up to terms that give a null contribution to $\sigu$
and $\sigd$, unless we include in $V_{\rm eff}$ also diagrams that do
not depend on the Higgs fields (such as, e.g., the one--loop diagram
involving gluinos and the two--loop diagram involving gluinos and
gluons).

Compact analytical formulae for the derivatives of the renormalized
effective potential in the gaugeless limit can be obtained with a
procedure similar to that of Ref.~\cite{noi}. 
The relevant field--dependent quantities are the top mass $m_t$, 
the stop masses $\msqu$ and $\msqd$, and the stop mixing angle $\ths \,$
(the top and stop phases $\varphi$ and $\widetilde{\varphi}$,
introduced in~\cite{noi} to take into account the dependence on the
CP--odd part of the Higgs fields, do not enter the computation of
$\sigu$ and $\sigd$). At the minimum of the effective potential,
the parameters in the stop sector are related by:
\be
\label{sin2t}
\sin 2 \ths = \frac{2\,\mt\,(A_t + \mu\,\cot\beta)}{\diff}\,,
\ee
where $A_t$ is the soft SUSY--breaking trilinear coupling of the stops
[notice that Eq.~(\ref{sin2t}) defines our convention for the sign of
$\mu\,$]. We will use in the following the shortcuts $\C2t \equiv \cos
2\ths$ and $\s2t \equiv \sin 2 \ths\,$. After a straightforward
application of the chain rule for the derivatives of the effective
potential, we get:
\bea
\label{sigu}
v_1^2\,\sigu & = & m_t\, \mu\, \cot\beta\, \s2t\, F \, ,\\
\label{sigd}
v_2^2\,\sigd & = & m_t\, A_t\, \s2t\, F + 2\, m_t^2\, G \, .
\eea 
The functions $F$ and $G$ are combinations of the derivatives of
$\Delta V$ with respect to the field--dependent parameters, computed
at the minimum of the \EP:
\bea
\label{deff}
F & = & \frac{\partial \Delta V}{\partial \msqu}
- \frac{\partial \Delta V}{\partial \msqd}
-\frac{2\, \C2t}{\msqu-\msqd} \,\frac{\partial \Delta V}{\partial \C2t}\, ,\\
&&\nn\\
\label{defg}
G & = & \frac{\partial \Delta V}{\partial m_t^2}
+ \frac{\partial \Delta V}{\partial \msqu}
+ \frac{\partial \Delta V}{\partial \msqd}\, .
\eea
The one--loop parts of $F$ and $G$, giving rise to $\oat$
contributions to $\sigu$ and $\sigd$, are easily computed from the
derivatives of $V_{1\ell}^{(1)}$.  In units of $N_c/(16\,\pi^2)$,
where $N_c=3$ is a color factor, they read:
\bea
\label{foneloop} 
F^{1\ell} & = & 
\msqu \left(\ln\frac{\msqu}{Q^2}-1\right) -
\msqd \left(\ln\frac{\msqd}{Q^2}-1\right) \, ,\\
&\nn\\ 
\label{goneloop} 
G^{1\ell} & = & 
\msqu \left(\ln\frac{\msqu}{Q^2}-1\right)
+\msqd \left(\ln\frac{\msqd}{Q^2}-1\right) -2\,
m_t^2 \left(\ln\frac{m_t^2}{Q^2}-1\right) \, .
\eea 
Although a naive, brute--force computation of the derivatives of the
renormalized two--loop potential $\widehat{V}_{2\ell}$ presents no
major conceptual difficulties, the number of terms involved blows up
quickly, giving rise to very long and complicated analytical
expressions. However, in the spirit of Ref.~\cite{davtausk}, it is
possible to obtain recursive relations for the derivatives of the
integral $I(m_1^2,m_2^2,m_3^2)$ with respect to the internal masses
(see the appendix A), that simplify considerably the results. In this
way, we obtained compact analytical formulae for the two--loop parts
of $F$ and $G$, giving rise to $\oatasatq$ contributions to $\sigu$
and $\sigd$. As a non trivial check of the correctness of our
computation, we have verified that the quantities $\mu^2$ and $m_3^2$
defined in Eqs.~(\ref{eqmu})--(\ref{eqm3}) obey the appropriate
two--loop RGE \cite{rge2l}, specialized to the gaugeless limit:
\bea
\label{rgemu}
\frac{\partial \mu^2}{\partial \ln Q^2} &=&
\frac{3\,\at}{4\,\pi} \,\mu^2 + \frac{\at\,\as}{\pi^2}\,\mu^2  
- \frac{9\,\at^2}{16\,\pi^2}\,\mu^2  \,,\\
&&\nn\\
\frac{\partial m_3^2}{\partial \ln Q^2} &=&
\frac{3\,\at}{4\,\pi} \left(\frac{m_3^2}{2} + A_t\,\mu\right)
+\frac{\at\,\as}{\pi^2} \left(\frac{m_3^2}{2} + A_t\,\mu -\mgl\,\mu\right)
-\frac{9\,\at^2}{16\,\pi^2} \left(\frac{m_3^2}{2} + 2\,A_t\,\mu\right)\,.\nn\\
\label{rgem3}
\eea
We have also checked explicitly that, in the special supersymmetric
limit in which all the soft--SUSY breaking parameters as well as $\mu$
are set to zero, such that $m_{\tilde{t}_1} = m_{\tilde{t}_2} = \mt$,
the two--loop parts of $F$ and $G$ are indeed vanishing.

For illustrative purposes, we present in the appendix B the explicit
formulae for the $\oatas$ corrections, valid for arbitrary values of
the relevant MSSM parameters. The corresponding formulae for the
$\oatq$ corrections are indeed rather long, thus we make them
available, upon request, in the form of a Fortran  code.

The computation described above allows us to obtain also the two--loop
$\oabas$ corrections~\footnote{In contrast, the $\oabqatab$
corrections would require a dedicated computation.} induced by the
bottom/sbottom sector ($h_b \equiv \sqrt{4 \pi \ab}$ being the bottom
Yukawa coupling), that can be relevant for large values of
$\tan\beta$. To this purpose, the substitutions $t \rightarrow b\,,\;
v_1 \leftrightarrow v_2$ (i.e., $\tan\beta \leftrightarrow \cot\beta$)
and $\sigu \leftrightarrow \sigd$ must be performed in the
corresponding formulae for the $\oatas$ part of the corrections. The
complications relative to the on--shell definition of the sbottom
parameters, discussed in Ref.~\cite{noibot}, do not arise in this case
since we are working in the $\drbar$ renormalization scheme. However,
the $\tan\beta$--enhanced threshold corrections \cite{threshbot} to
the relation between $h_b$ and the bottom mass $m_b$ must be resummed
to all orders~\cite{resumbot} in a redefinition of the bottom Yukawa
coupling (see e.g. Ref.~\cite{noibot} for the details).


\section{Numerical results}
\label{sec:numerical}

In this section we discuss the numerical effect of our $\oatasatq$
two--loop corrections on the minimization conditions of the MSSM
effective potential.

For definiteness, we work in the mSUGRA scenario, in which the MSSM
Lagrangian at the large scale $\mgut$ contains only five independent
mass parameters: a common soft SUSY-breaking scalar mass $\m0$, a
common soft gaugino mass $\mhf$, a common soft trilinear term $A_0$,
the superpotential Higgs mixing parameter $\mu_0$ and its soft
SUSY-breaking counterpart $B_0$ (the subscript ``0'' denotes the fact
that the parameters are computed at the boundary scale). The soft
Higgs mixing parameter $B$ has the dimensions of a squared mass, and
is defined in such a way that in the low--energy Higgs potential of
Eq.~(\ref{V0}) it coincides with $m_3^2$ (to avoid confusion, we will
refer to $B$ as to $m_3^2\,$ from now on).  As anticipated in section
\ref{sec:general}, rather than providing input values for all the five
mass parameters at the GUT scale, we assume that the electroweak
symmetry is successfully broken at the weak scale, and we trade
$\mu_0$ and $m_3^2(\mgut)$ for the weak scale input parameters $v$ and
$\tan\beta$.

Before discussing our results, it is useful to describe in some detail
the numerical procedure for the renormalization group evolution of the
MSSM parameters. We start by defining the $SU(3)\times SU(2)_L\times
U(1)_Y$ gauge couplings at the weak scale (which we identify with the
pole $Z$ boson mass, $\polemz= 91.187$ GeV), from the running weak
mixing angle $\hat{s}^{\,2} = 0.2315$, the electromagnetic coupling
$\widehat{\alpha}_{\scriptscriptstyle EM} = 1/127.9$ and the strong
coupling $\as = 0.119$. The electroweak symmetry breaking parameter,
$v^2 = v_1^2 + v_2^2\,$, is defined in terms the muon decay constant
according to the relation $v = (\sq2\,G_{\mu})^{-1/2}=246.218$ GeV,
and then translated to the $\drbar$ scheme by means of the formulae of
Ref.~\cite{noi}. In addition, $\tan\beta = v_2/v_1$ is taken as an
input parameter at the weak scale, allowing us to determine $v_1$ and
$v_2$. The Yukawa couplings of the light SM fermions are obtained from
the corresponding masses at the scale $Q=2$ GeV, and then evolved up
to $\polemz$ by means of the two--loop SM RGEs~\cite{Arason}. 
In the case of the
bottom coupling, the $\tan\beta$--enhanced threshold corrections
\cite{threshbot,resumbot} to the relation between $h_b$ and $m_b$ are
included at the scale $\polemz$ according to the formulae of
Ref.~\cite{noibot}. Finally, the top Yukawa coupling is also defined
at the scale $\polemz$ through the relation $h_t = \sq2\,m_t/v_2$,
where $m_t$ is the $\drbar$ mass for the top quark, obtained from the
pole mass $M_t= 174.3$ GeV by means of the formulae of
Ref.~\cite{noi}.

The evolution of the MSSM parameters from the weak scale to the GUT
scale, and back, is performed by means of the full one--loop MSSM
RGEs.  However, for consistency with our two--loop $\oatasatq$
analysis of the REWSB conditions, we supplement the RGEs with the
two--loop strong and top--Yukawa contributions as given in
Ref.~\cite{rge2l}. We also make use of the two--loop RGEs in the 
Landau gauge for the VEVs, following Refs.~\cite{martinEP,Yamada}.
As a first step, we evolve the gauge and Yukawa couplings from the
scale $\polemz$ up to the scale where the $SU(2)_L\times U(1)_Y$ gauge
couplings $g_1$ and $g_2$ meet, that we identify with $\mgut$ (we do
not force the strong gauge coupling $g_3$ to meet  $g_1$ and $g_2$).  
At this scale, which turns out to be of the order of $10^{16}$ GeV, we
set the input boundary conditions for the soft SUSY--breaking masses
$\m0,\,\mhf$ and $A_0$.  The values of $\mu_0$ and $m_3^2(\mgut)$, to
be later determined from the REWSB conditions, are provisionally set
to zero.
At this point, we start an iterative procedure: first we run the MSSM
parameters from $\mgut$ down to some scale $\qmin$, of the order of
the weak scale, where the values of $\mu^2(\qmin)$ and $m_3^2(\qmin)$
are computed through Eqs.~(\ref{eqmu}) and (\ref{eqm3}), with the sign
of $\mu$ supplied as an extra input parameter. Then we run all the
parameters, including $\mu$ and $m_3^2$, down to the scale $\polemz$,
which we regard as the end point of the RGE evolution. 
At this scale, we compute the threshold corrections
to the top and bottom Yukawa couplings, using the newly obtained values
of the relevant MSSM parameters.
Finally, we run
all the parameters back to the scale $\mgut$, where the resulting values
for $\mu$ and $m_3^2$ are taken as new guesses for the corresponding
boundary conditions. We iterate the procedure until convergence is
reached, i.e.~the values of $\mu^2(\qmin)$ and $m_3^2(\qmin)$ obtained
from the RGE evolution of $\mu_0$ and $m_3^2(\mgut)$ coincide with
those obtained from the minimization conditions (\ref{eqmu}) and
(\ref{eqm3}). If however $\mu^2$ turns out to be negative, then our
choice of input parameters is inconsistent (i.e., it does not lead to
successful REWSB) and must be discarded.

\begin{figure}[p]
\begin{center}
\epsfig{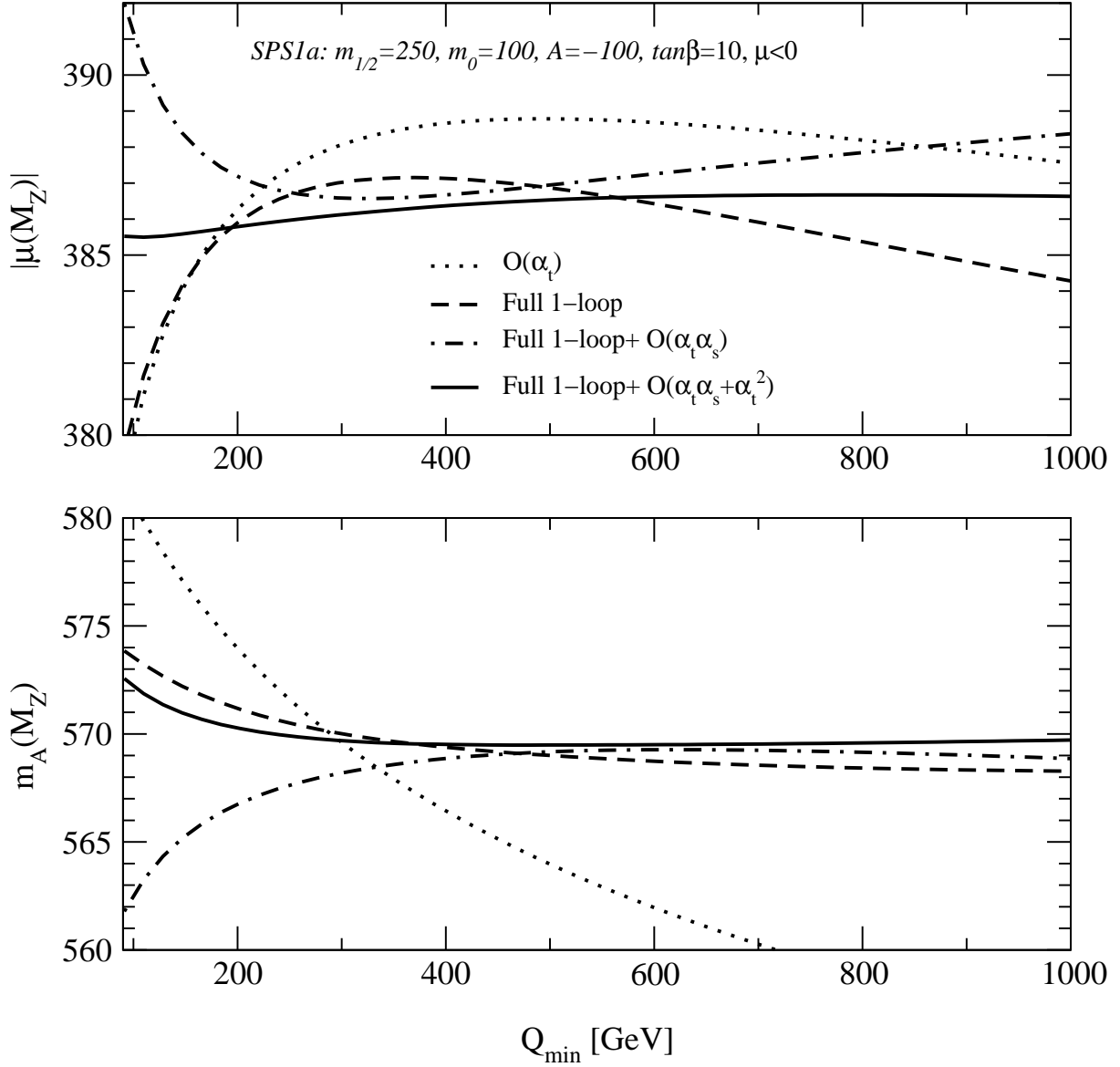}
\end{center}
\caption{The $\drbar$ parameters $|\mu(\polemz)|$ (upper plot) and
$\ma(\polemz)$ (lower plot) as a function of the scale $\qmin$ at
which the minimization conditions of the effective potential are
imposed. The input parameters of the mSUGRA scenario are chosen as in
the Snowmass Point SPS 1a \cite{snowmass}. The meaning of the different
curves is shown in the caption and explained in the text. The mass
of the top quark is taken to be 174.3 GeV.}
\label{fig:sps1a}
\end{figure}

\begin{figure}[p]
\begin{center}
\epsfig{figure=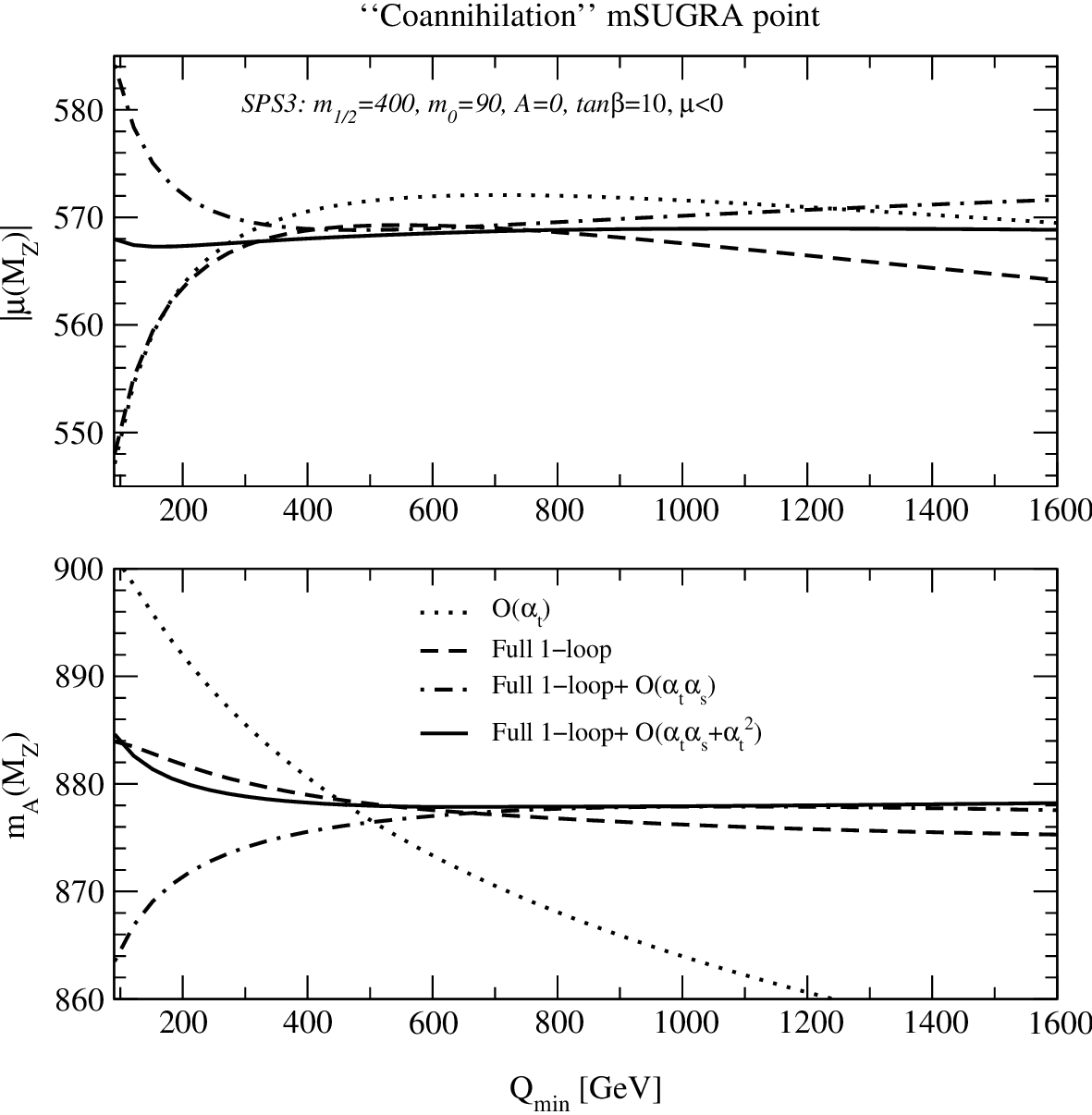,width=16cm}
\end{center}
\caption{Same as Fig.~\ref{fig:sps1a} for the 
Snowmass Point SPS 3.}
\label{fig:sps3}
\end{figure}

\begin{figure}[p]
\begin{center}
\epsfig{figure=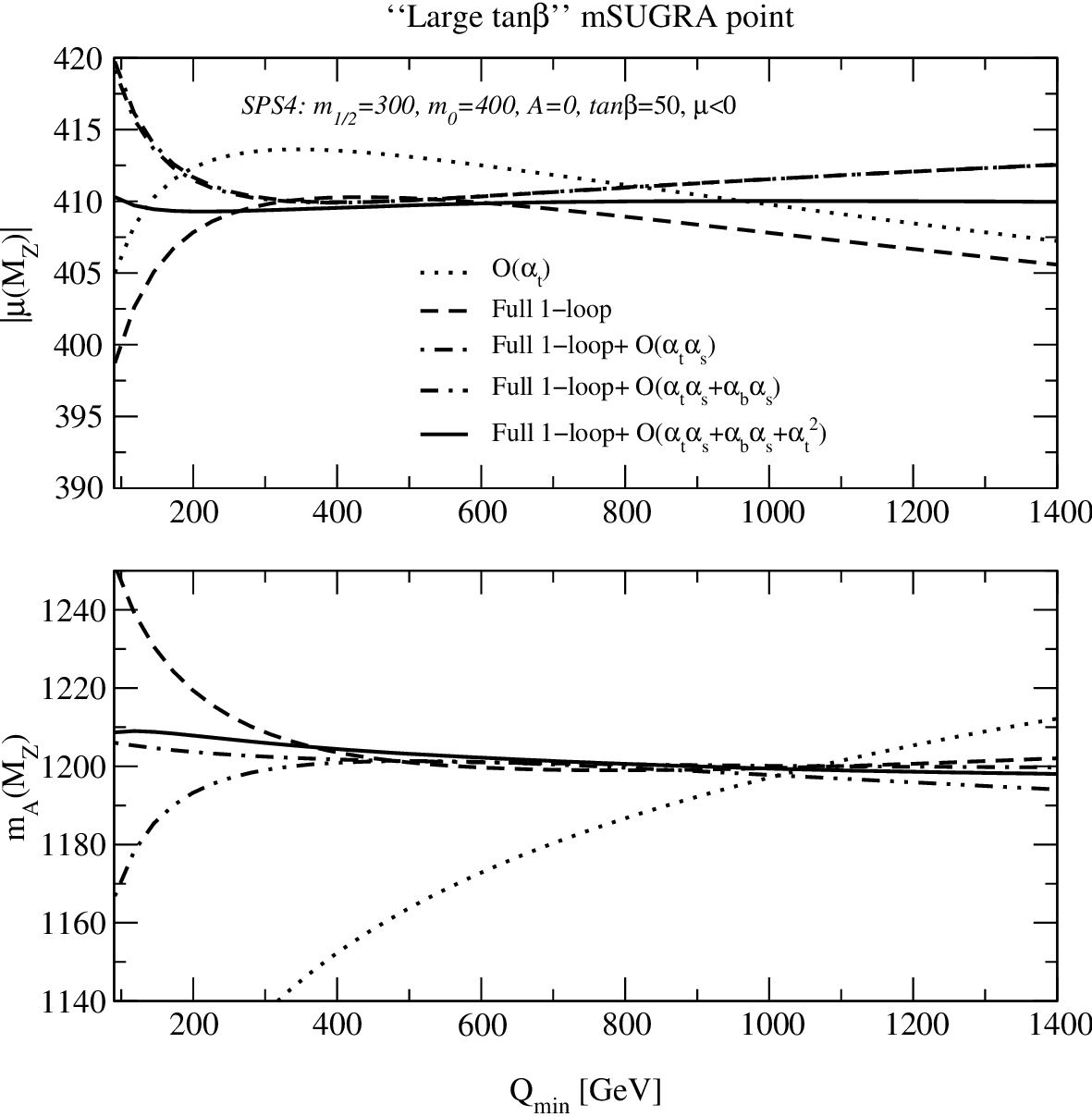,width=16cm}
\end{center}
\caption{Same as Fig.~\ref{fig:sps1a} for the 
Snowmass Point SPS 4.}
\label{fig:sps4}
\end{figure}

\begin{figure}[p]
\begin{center}
\epsfig{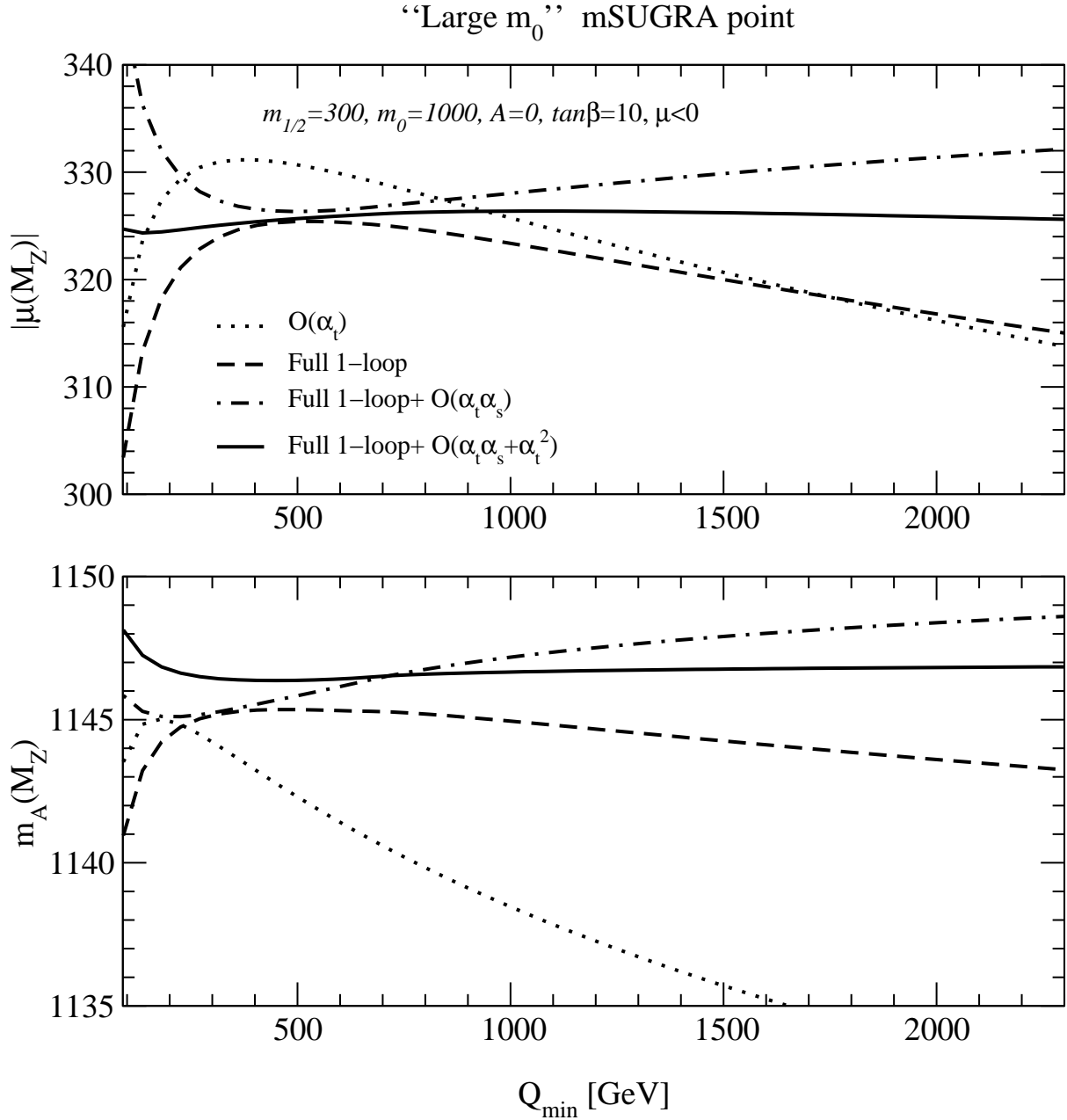}
\end{center}
\caption{Same as Fig.~\ref{fig:sps1a} for the ``Large $\m0$''
mSUGRA point proposed in section~\ref{sec:numerical}.}
\label{fig:largem0}
\end{figure}

In order to discuss the effect of our $\oatasatq$ corrections to the
REWSB conditions, we show in Figs.~\ref{fig:sps1a}--\ref{fig:largem0} the
values of $|\mu(\polemz)|$ and $\ma(\polemz)$, the latter obtained
through $\ma^2(\polemz) = - 2\,m_3^2(\polemz)/ \sin 2\beta\,$, as
functions of the minimization scale $\qmin$. Stability of the results
with respect to moderate changes in $\qmin$ (which should anyway lie
in the weak range, i.e.~between $\polemz$ and a few TeV) indicates
that the higher--order corrections not included in the computation of
$\mu$ and $\ma$ are small and can be safely neglected. We will see
that in general the inclusion of our $\oatasatq$ corrections
significantly improves the scale--dependence of the results.

In the choice of the input parameters, we refer to the so--called
Snowmass Points \cite{snowmass}, which represent typical ``benchmark''
scenarios that are commonly investigated in the phenomenological
analyses of the mSUGRA parameter space. In particular,
Figs.~\ref{fig:sps1a}--\ref{fig:sps5} correspond to:
\[
\begin{array}{lccccc}
\vspace*{2mm} 
{\rm SPS\; 1a}: &  
\m0 = 100\; \gev, &
\mhf = 250\; \gev, &
A_0 = -100\; \gev, &
\tan\beta = 10\,, & 
\mu < 0\,, \\
\vspace*{2mm} 
{\rm SPS\; 3}: & 
\m0 = 90\; \gev, &
\mhf = 400\; \gev, &
A_0 = 0\,, &
\tan\beta = 10\,, &
\mu < 0\,,\\ 
\vspace*{2mm} 
{\rm SPS\; 4}: & 
\m0 = 400\; \gev, &
\mhf = 300\; \gev, &
A_0 = 0\,, &
\tan\beta = 50\,, &
\mu < 0\,,\\ 
\vspace*{2mm} 
{\rm SPS\; 5}: & 
\m0 = 150\; \gev, &
\mhf = 300\; \gev, &
A_0 = -1\; {\rm TeV}\,, &
\tan\beta = 5\,, &
\mu < 0\,,
\end{array}
\]
respectively. Notice that our convention for the sign of $\mu$ differs
from the one in Ref.~\cite{snowmass}, where a discussion on the
characteristics of the various scenarios can be found. A further
scenario, denoted as ``Focus point'' (SPS2) and characterized by a
common scalar mass, $\m0 = 1450$ GeV, much larger than the common
fermion mass, $\mhf = 300$ GeV, has also been suggested in
Ref.~\cite{snowmass}. However, we found that in this scenario the
results for $\mu$ and $\ma$, including the qualitative effect of the
various corrections and the occurrence of REWSB itself, depend
dramatically on very small adjustments of the input value for the top
pole mass (e.g.~$|\mu(\polemz)|$ varies roughly between 400 and 100
GeV if $M_t$ is varied between 174.3 and 175 GeV). The extreme
sensitivity of the SPS2 scenario on the input top mass has already
been discussed in Ref.~\cite{kraml}. Since this scenario appears to
lead to unstable results, we will not consider it further in this
work.  However, in order to investigate the situation in which the
common scalar mass is considerably larger than the common fermion
mass, we show in Fig.~\ref{fig:largem0} a scenario with $\m0=1$ TeV
and $\mhf = 300$ GeV (the other parameters being chosen as $A_0=0\,,\,
\tan\beta = 10$ and $\mu <0$). We have checked that this ``Large $m_0$''
 scenario is
not unreasonably sensitive to small variations in the input top mass.

In all the plots of Figs.~\ref{fig:sps1a}--\ref{fig:largem0} the
minimization scale varies in the range $\polemz < \qmin < 2\,\qbest$,
where $\qbest \equiv (\m0^2+ 4\,\mhf^2)^{1/2}$ is a scale roughly
comparable with the squark masses.  The dotted curves in the upper and
lower panels of each figure represent $|\mu(\polemz)|$ and
$\ma(\polemz)$, respectively, as obtained by including in the
minimization conditions only the one--loop $\oat$ top/stop
contributions to $\sigu$ and $\sigd$; the dashed lines include instead
the full one--loop computation of $\sigu$ and $\sigd$; the dot--dashed
lines include in addition the two--loop $\oatas$ contributions;
finally, the solid lines include our full two--loop result, i.e.~the
$\oatasatq$ contributions to $\sigu$ and $\sigd$. In
Fig.~\ref{fig:sps4}, corresponding to the ``large $\tan\beta$'' (SPS4)
scenario with $\tan\beta = 50$, we show also the effect of the
$\oabas$ corrections, obtained from the $\oatas$ ones as described at
the end of the previous section.  The dot--dot--dashed lines in
Fig.~\ref{fig:sps4} include the $\oatasabas$ contributions, while the
solid lines represent the full $\oatasabasatq$ result. The effect of
the $\oabas$ corrections is indeed negligible in the other SPS
scenarios, where $\tan\beta$ takes on more moderate values.  In any
case we include the $\oabas$ corrections in all the scenarios we
investigate here.

We see from Figs.~\ref{fig:sps1a}--\ref{fig:largem0} that the
inclusion of the two--loop $\oatasatq$ corrections improves the
dependence of $|\mu(\polemz)|$ and $\ma(\polemz)$ on the minimization
scale with respect to the full one--loop result. The effect is
particularly manifest in the case of $\mu$, where, in all the
scenarios, the two--loop corrected result appears to depend only very
weakly (within 3 GeV at most) on $\qmin$, while the one--loop result
tends to decrease for both small and large $\qmin$. In the case of
$\ma$, the improvement is less striking: although the two--loop
corrected result has in general a better scale dependence than its
one--loop counterpart, especially for increasing values of $\qmin$, a
small residual scale dependence is visible in most plots when $\qmin$
gets close to $\polemz$. In any scenario the residual uncertainty on
$\ma$ is never larger than 10 GeV (the latter case occurring in the
``large $\tan\beta$'' scenario, where $\ma(\polemz)$ is around 1200
GeV) and might be due to the corrections that we neglect in our
two--loop computation of the tadpoles, i.e.~those controlled by the
electroweak gauge couplings and, for large $\tan\beta$, those of
$\oabqatab\,$.
From the small residual scale dependence visible in
Figs.~\ref{fig:sps1a}--\ref{fig:largem0}, we estimate that the effect
of the neglected two--loop and higher--order corrections on 
the parameters $\mu$ and $\ma$ should be at most of 1\%. 

Other interesting observations can be drawn from 
Figs.~\ref{fig:sps1a}--\ref{fig:largem0}: first of all, 
at the one--loop level,
the inclusion of the top/stop contributions only is in general not a
good approximation of the full result (this has been already observed
e.g.~in Ref.~\cite{spanos}).  Moreover, at the two--loop level, a
significant compensation occurs between the $\oatas$ and the $\oatq$
contributions to the tadpoles, thus including only the former may lead
to rather inaccurate predictions (as shown by the dot--dashed curves).
This partial cancellation between the $\oatas$ and $\oatq$ corrections
is similar to the one occurring in the case of the Higgs
masses~\cite{hh,ez,noi}. Finally, it is worth noticing that in the
``large $\tan\beta$'' scenario of Fig.~\ref{fig:sps4} the inclusion of
the $\oabas$ corrections to the tadpoles has a sizeable effect on the
minimization scale dependence of $\ma(\polemz)$ (see the difference
between the dot--dashed and dot--dot--dashed curves), while it does
not affect significantly that of $|\mu(\polemz)|$.

It is clear from the above discussion that the numerical effect on
$|\mu(\polemz)|$ and $\ma(\polemz)$ of the two--loop corrections to
$\sigu$ and $\sigd$ depends critically on the choice of the
minimization scale. In all the plots we find a range of values of
$\qmin$, usually in the vicinity of $\qbest$, for which the one--loop
and two--loop curves are close to each other, implying that the effect
of the $\oatasatq$ corrections is small. On the other hand, for values
of $\qmin$ far from this optimal choice, the omission of the two--loop
corrections can lead to an error of several (possibly, tenths of) GeV,
especially in the case of $\mu$. Thus, the proper inclusion the
$\oatasatq$ two--loop tadpole corrections on the top of the full
one--loop ones allows us to obtain more precise and reliable results
for $\mu$ and $\ma$, i.e.~results that do not depend on a preconceived
choice of the minimization scale for the MSSM effective potential.
 

\section{Conclusions}

In this paper we presented explicit and general results for the
$\oatasatq$ two--loop corrections to the minimization conditions of
the MSSM effective potential, which translate into corrections to the
$\drbar$--renormalized parameters $\mu$ and $\ma$. We discussed the
numerical impact of our corrections in some representative scenarios
of gravity--mediated SUSY breaking, and we found that the inclusion of
the $\oatasatq$ corrections significantly improves the renormalization
scale dependence of the results. Due to partial cancellations between
the $\oatas$ and $\oatq$ corrections, including only the former may
lead to inaccurate results. Our corrections are also required for
consistency in the $\oatasatq$ two--loop computations of the MSSM
Higgs masses, if the parameters entering the tree--level Higgs mass
matrix are computed via renormalization group evolution from a set of
high energy boundary conditions.

A complete study of the electroweak symmetry breaking at the two--loop
level would require also the knowledge of the corrections that are
neglected in our gaugeless limit, among which the most relevant are
controlled by the bottom Yukawa coupling and the electroweak gauge
couplings. Concerning the corrections controlled by $\ab$, they can be
numerically non--negligible only for large values of $\tan\beta$.  As
discussed at the end of section~\ref{sec:twoloop}, the formulae for
the $\oabas$ corrections can be obtained by performing simple
substitutions in their $\oatas$ counterparts. On the other hand, the
$\oabqatab$ corrections, which in some cases might be as relevant as
the $\oabas$ ones, cannot be obtained in a straightforward way from
the presently computed corrections and would require further work.

In Ref.~\cite{martinEP} a complete two--loop computation of the MSSM
\EP\ is presented, including also the terms controlled by the
electroweak gauge couplings that are neglected in our analysis. The
inclusion of such terms improves further the scale dependence of the 
parameters $\mu$ and $\ma$, that in Ref.~\cite{martinEP} are determined
through a numerical minimization of the effective potential. 
However, the explicit
analytical formulae for the two--loop tadpoles $\sigu$ and $\sigd$,
which are usually needed for practical applications, would be quite
involved in the general case, and have not been presented so far.

In conclusion, our work should lead to a more precise and reliable
determination of the MSSM parameters at the weak scale, once the
boundary conditions are provided at some larger scale according to the
underlying theory of SUSY breaking.  The fact that, among its many
attractive features, the MSSM provides a natural mechanism for
breaking radiatively the electroweak symmetry, with the heavy top
quark mass nicely falling in the required 
range\footnote{As an example, we find that in the ``typical''  mSUGRA (SPS1a)
scenario the acceptable range for the top quark mass is 
$80~ {\rm GeV} < M_t < 215~{\rm GeV}$.}, seems to indicate the
MSSM as the most viable theory for physics at the weak scale. However, only
the forthcoming experimental results from the Tevatron and the  LHC
will tell us if this is indeed the case.

\newpage

\section*{Acknowledgments}

We thank G.~Degrassi, S.~P.~Martin, W.~Porod, 
K.~Tamvakis, F.~Zwirner and especially M.~Drees for useful comments and
discussions. A.~D.~also thanks J.~R.~Espinosa and R.~J.~Zhang for
discussions in the early stages of the project. This work was
partially carried out at the Physics Department of the University of
Bonn, and it was partially supported by the European Programmes
HPRN-CT-2000-00148 (Across the Energy Frontier) and HPRN-CT-2000-00149
(Collider Physics).


\section*{Appendix A: Two--loop integrals}
\begin{appendletterA}

We give here explicit expressions for the momentum integrals that
appear in the two-loop part of the \EP. The basic integrals in $d = 4
- 2\epsilon$ dimensions are:
\bea
\label{def2J}
\hspace{-1cm}
\frac{1\;\;}{(16 \,\pi^2)^2}\, J(x,y)
& \equiv &
- \frac{\mu^{2(4-d)}}{(2 \pi)^{2d}}\,
\int\!\int\,\frac{{\rm d}^d p\,{\rm d}^d q}{[p^2-x][q^2-y]}\,,\\ 
&&\nn\\
\label{def2I}
\hspace{-1cm}
\frac{1}{(16 \,\pi^2)^2}\,I(x,y,z)
& \equiv &
\frac{\mu^{2(4-d)}}{(2 \pi)^{2d}}\,
\int\!\int\,\frac{{\rm d}^d p\,{\rm d}^d q}
{[p^2-x][q^2-y][(p-q)^2-z]}\,.\\ 
&&\nn
\eea
Following Ref.~\cite{davtausk}, the functions $J(x,y)$ 
and $I(x,y,z)$ defined in Eqs.~(\ref{def2J})--(\ref{def2I}) are: 
\bea
\label{expJ}
J(x,y) & = & \frac{x y}{\epsilon^2} 
- \frac{x y}{\epsilon}\, \left( \lnb x + \lnb y -2 \right)
-x y\,\left[ 2\,\lnb x + 2\,\lnb y - \hlf\,\lnb^{\,2} xy 
- \left( 3 + \frac{\pi^2}{6} \right)\,\right]\,,\\
&&\nn\\
I(x,y,z) & = &
-  \frac{x+ y+z}{2 \epsilon^2}
\;+\; \frac{1}{\epsilon} \left[x \,\lnb x + y \,\lnb y + z \, \lnb z
-\frac{3}{2}\,(x+y+z)\right]\nn\\
&& + \hlf\,(x\,\lnb y\,\lnb z + y\,\lnb x\,\lnb z + z\,\lnb x\,\lnb y )
- \frac{x+y+z}{2}\,(7 + \pi^2/6)\nn\\
&& - \hlf\,(x \,\lnb x + y \,\lnb y + z \, \lnb z)
\,(\lnb x + \lnb y + \lnb z -6)
- \frac{\Delta(x,y,z)}{2 z} \,\Phi(x,y,z)\,.
\eea

\noindent
In the above formulae, $\lnb x$ stands for $\ln(x/Q^2)$, where $Q^2 =
4 \pi \mu^2 e^{-\gamma_E}$ is the renormalization scale ($\gamma_E$ is
the Euler constant). The functions $\Delta$ and $\Phi$ are respectively:
\bea
\label{defdelta}
\Delta(x,y,z) & = & x^2 + y^2 + z^2 - 2\,(x y + x z + y z)\, ,\\
&&\nn\\
\label{defphi}
\Phi\,(x,y,z) & = & \frac{1}{\lambda} \left[
2\,\ln x_+\,\ln x_- - \ln u \,\ln v -
2\, \biggr( {\rm Li}_2 (x_+) + {\rm Li}_2 (x_-) \biggr) + 
\frac{\pi^2}{3} \right] \, ,
\eea
where ${\rm Li}_2(z) = -\int_0^z {\rm d}t \left[\ln(1-t)/t\right]$ 
is the dilogarithm function and the auxiliary (complex) variables are:
\be
u = \frac{x}{z}\,,\;\;\;\;\;\;
v = \frac{y}{z}\,,\;\;\;\;\;\;
\lambda = \sqrt{(1-u-v)^2 - 4 \,u\, v}\,,\;\;\;\;\;\;
x_{\pm} = \frac{1}{2}\,\left[1 \pm (u-v) - \lambda\right] \, .
\ee
The definition (\ref{defphi}) is valid for the case $x/z < 1$ and 
$y/z < 1$. The other branches of $\Phi$ can be obtained using the symmetry 
properties:
\be
\label{symm}
\Phi\,(x,y,z) = \Phi\,(y,x,z)\,,\hspace{1cm}
x\,\Phi\,(x,y,z) = z \, \Phi\,(z,y,x) \, .
\ee
Finally, the following recursive relation for the derivatives of
$\Phi$ proved very useful~\footnote{We thank G.~Degrassi for
explanations on how to derive Eq.~(\ref{dphidx}).} for obtaining
compact analytical results:
\be
\label{dphidx}
\Delta(x,y,z)\,\frac{\partial \,\Phi(x,y,z)}{\partial\, x} 
= (y+z-x)\,\Phi(x,y,z) + 
\frac{z}{x}\,\left[ (y-z)\,\ln\frac{z}{y}
+ x\,\left( \ln\frac{x}{y} + \ln\frac{x}{z}\right)\,\right]\, .
\ee
The derivatives of $\Phi$ with respect to $y$ and $z$ can be obtained
from the above equation with the help of the symmetry properties of 
Eq.~(\ref{symm}).

\end{appendletterA}

\vspace{0.4cm}
\section*{Appendix B: Explicit formulae for the $\oatas$ corrections}
\begin{appendletterB}
\vspace{0.1cm}

We present here explicit expressions for the two--loop part of the
functions $F$ and $G$, giving rise to $\oatas$ corrections to $\sigu$
and $\sigd$. These formulae are valid when the parameters entering the
one--loop parts of $F$ and $G$ are expressed in the $\drbar$ scheme
and computed at the renormalization scale $Q$. In units of
$g_s^2\,C_F\,N_c/(16\pi^2)^2$, where $C_F = 4/3$ and $N_c = 3$ are
color factors, $F^{2\ell}$ and $G^{2\ell}$ read:

\bea
\label{ftwoloop}
F^{2\ell} & = & 
\frac{4\,\mgl\,\mt}{\s2t}\,(1+4\,\C2t^2)
-\left[2\,(\diff) + \frac{4\,\mgl\,\mt}{\s2t}\right]\,\mylg\,\mylt\nn\\
&&\nn\\
&&
- 2\,(4-\s2t^2)\,(\diff)
+ \frac{4\,\msqu\,\msqd - \s2t^2\,(\msqu+\msqd)^2}{\diff}
\,\myltu\,\myltd\nn\\
&&\nn\\
&+& \hspace{-0.2cm}\left\{\hspace{0.3cm}
\left[4\,(\mgl^2+\mt^2+2\,\msqu)-\s2t^2\,(3\,\msqu+\msqd)
-\frac{16\,\C2t^2\,\mgl\,\mt\,\msqu}{\s2t\,(\diff)} -4\,\s2t\,\mgl\,\mt
\right]\,\myltu \right.\nn\\
&&\nn\\
&&+\frac{\msqu}{\diff}\,\left[
\s2t^2\,(\msqu+\msqd)-2\,(2\,\msqu-\msqd)\right]\,\myltuq\nn\\
&&\nn\\
&&+ 2\,\left[\msqu-\mgl^2-\mt^2+\mgl\,\mt\,\s2t
+\frac{2\,\C2t^2\,\mgl\,\mt\,\msqu}{\s2t\,(\diff)}\right]\,
\ln\frac{\mgl^2\,\mt^2}{Q^4}\,\myltu\nn\\
&&\nn\\
&&+\frac{4\,\mgl\,\mt\,\C2t^2\,(\mt^2-\mgl^2)}{\s2t\,(\diff)}\,
\ln\frac{\mt^2}{\mgl^2}\,\myltu
+\left[ \frac{4\,\mgl^2\,\mt^2+ 2\,\Delta(\mgl^2,\mt^2,\msqu)}{\msqu}
\right.\nn\\
&&\nn\\
&&\left.
-\frac{2\,\mgl\,\mt\,\s2t}{\msqu}\,(\mgl^2+\mt^2-\msqu)
+\frac{4\,\C2t^2\,\mgl\,\mt\,\Delta(\mgl^2,\mt^2,\msqu)}
{\s2t\,\msqu\,(\diff)}\,\right]\,\Phi(\mgl^2,\mt^2,\msqu)\nn\\
&&\nn\\
&& \left. \hspace{5cm}\phantom{\myltuq}
- (\msqu \leftrightarrow \msqd\,, \;\s2t \rightarrow -\s2t)
\;\;\right\}\,,\\
&&\nn\\
&&\nn\\
&&\nn\\
\label{gtwoloop}
G^{2\ell} & = & 
\frac{5\,\mgl\,\s2t}{\mt}\,(\diff)
-10\, (\msqu+\msqd-2\,\mt^2) - 4\,\mgl^2 
+ 12\,\mt^2\,\left(\myltq - 2 \mylt\right)\nn\\
&&\nn\\
&& +\left[ 4\,\mgl^2 - \frac{\mgl\,\s2t}{\mt}\,(\diff)\right]
\,\mylg\,\mylt
+ \s2t^2\,(\msqu+\msqd)\,\myltu\,\myltd \nn\\
&&\nn\\
&+& \hspace{-0.2cm}\left\{\hspace{0.3cm}
\left[4\,(\mgl^2+\mt^2+2\,\msqu) + \s2t^2\,(\diff)
- \frac{4\,\mgl\,\s2t}{\mt}\,(\mt^2+\msqu)\right]\,\myltu \right.\nn\\
&&\nn\\
&& +\left[\frac{\mgl\,\s2t}{\mt}\,(5\,\mt^2-\mgl^2+\msqu)
-2\,(\mgl^2+2\,\mt^2)\right]\,\mylt\,\myltu\nn\\
&&\nn\\
&& +\left[\frac{\mgl\,\s2t}{\mt}\,(\mgl^2-\mt^2+\msqu)
-2\,\mgl^2\right]\,\mylg\,\myltu
-(2 + \s2t^2)\,\msqu\,\myltuq\nn\\
&&\nn\\
&& +\left[
2\,\frac{\mgl^2}{\msqu}\,
(\mgl^2 +\mt^2-\msqu-2\,\mgl\,\mt\,\s2t)
+\frac{\mgl\,\s2t}{\mt\,\msqu}\,\Delta(\mgl^2,\mt^2,\msqu)
\right]\,\Phi(\mgl^2,\mt^2,\msqu)\nn\\
&&\nn\\
&& \left. \hspace{5cm}\phantom{\myltuq}
+ (\msqu \leftrightarrow \msqd\,, \;\s2t \rightarrow -\s2t)
\;\;\right\}\, .
\eea
The functions $\Delta(x,y,z)$ and $\Phi(x,y,z)$ appearing in
$F^{2\ell}$ and $G^{2\ell}$ are defined in
Eqs.~(\ref{defdelta})--(\ref{defphi}).  The parameter $\s2t$ is
defined in Eq.~(\ref{sin2t}).

\end{appendletterB}


%

%
\begin{figure}[p]
\begin{center}
\epsfig{figure=SPS5.eps,width=16cm}
\end{center}
\caption{Same as Fig.~\ref{fig:sps1a} for the 
Snowmass Point SPS 5.}
\label{fig:sps5}
\end{figure}

\end{document}